\documentclass{article}

\usepackage{PRIMEarxiv}

\usepackage[utf8]{inputenc} 
\usepackage[T1]{fontenc}    
\usepackage{hyperref}       
\usepackage{url}            
\usepackage{booktabs}       
\usepackage{amsfonts}       
\usepackage{nicefrac}       
\usepackage{microtype}      
\usepackage{lipsum}
\usepackage{fancyhdr}       
\usepackage{graphicx}       
\graphicspath{{media/}}     

\usepackage{amsmath}
\usepackage{multicol}
\usepackage{multirow}
\usepackage{booktabs}
\usepackage[super]{nth}
\usepackage{colortbl}
\usepackage{caption}
\usepackage{subcaption}

\title{Large Language Models are unreliable for Cyber Threat Intelligence
\thanks{\textit{\underline{Citation}}: 
\textbf{Mezzi, E., Massacci, F., \& Tuma, K. (2025, August). Large language models are unreliable for cyber threat intelligence. In International Conference on Availability, Reliability and Security (pp. 343-364). Cham: Springer Nature Switzerland. Conference version DOI: \url{https://doi.org/10.1007/978-3-032-00627-1_17}.}} 
} 

\author{
  Emanuele Mezzi \\
  \texttt{e.mezzi@vu.nl} \\
   \And
  Fabio Massacci \\
  \texttt{fabio.massacci@ieee.org} \\
   \And
  Katja Tuma \\
  \texttt{k.tuma@tue.nl} \\
}

\begin{document}
\maketitle

\begin{abstract}
Several recent works have argued that Large Language Models (LLMs) can be used to tame the data deluge in the cybersecurity field, by improving the automation of Cyber Threat Intelligence (CTI) tasks. This work presents an evaluation methodology that other than allowing to test LLMs on CTI tasks when using zero-shot learning, few-shot learning and fine-tuning, also allows to quantify their consistency and their confidence level. We run experiments with three state-of-the-art LLMs and a dataset of 350 threat intelligence reports and present new evidence of potential security risks in relying on LLMs for CTI. We show how LLMs cannot guarantee sufficient performance on real-size reports while also being inconsistent and overconfident. Few-shot learning and fine-tuning only partially improve the results, thus posing doubts about the possibility of using LLMs for CTI scenarios, where labelled datasets are lacking and where confidence is a fundamental factor.
\end{abstract}


\section{Introduction}

The number of vulnerabilities is becoming overwhelming: De Smale et al. \cite{de2023no} report that companies have reduced the raw intake of vulnerability information by 95\%. In the quest to only do the work that really matters \cite{di2022software}, Cyber threat intelligence (CTI) seems to be the new coping strategy. Unfortunately, despite standardization efforts such as STIX \cite{barnum2012standardizing}, TAXII \cite{connolly2014trusted}, and MISP \cite{wagner2016misp}, CTI still requires humans to manage the massive amount of natural language information \cite{nainna2023factors}. 

Large Language Models (LLMs) seem to be the solution to tame the CTI data deluge \cite{saddi2024examine,sai2024generative} compared to pre-trained language models (PLM) such as BERT, used to automate the identification of Advanced Persistent Threat (APT) attack events \cite{chen2023automatically,xiang2023apt} and the extraction of knowledge graph (KG) \cite{wang2024knowcti}. Recent papers report high accuracy levels of LLMs processing CTI. Patsakis et al. \cite{Patsakis2024} report 89\% accuracy for extracting indicators of attack compromise and Hu et al. \cite{hu2024llm} used few-shot learning and fine-tuning in both named entity recognition and MITRE's Tactics, Techniques and Procedures classification, with precision of 88\% and 97\%, respectively. Prompt engineering \cite{chen2023unleashing} and model benchmarking for CTI tasks \cite{ji2024sevenllm} also produced promising results. CTI extraction is being augmented to use the LLM as a chatbot and requests to furnish information regarding a specific subject \cite{ferrag2024generative}. LLMs are known to hallucinate \cite{huang2023survey} but their CTI applications seem immune.

Unfortunately, \emph{such promises are not based on real CTI reports}. Table \ref{tab:datasets_length} shows the lengths of inputs employed to perform CTI extraction, and compares them to the CISA's emergency directive on SolarWinds. Previous claims analyzed sentences instead of reports or at best small paragraphs, shorter than the abstract of this paper. Hence our first question:

\begin{table}[t]
\begin{center}
\caption{LLMs extracting CTI info: papers vs reality.}
\begin{minipage}{\columnwidth}\footnotesize
LLM performance is strong when evaluated on inputs shorter than the abstract of this manuscript. However, in realistic scenarios involving full-length cybersecurity reports such as CISA's Emergency Directive 21-01 (SolarWinds) \cite{cisa_solarwinds_2021} or the 350 MITRE's APT reports analyzed in \cite{di2022software}, averaging 3,009 words LLM performance degrades significantly due to the increased input length and complexity.

\end{minipage}
\label{tab:datasets_length}
\resizebox{0.70\columnwidth}{!}{%
\begin{tabular}{llrr}
\toprule
\textbf{Dataset}                                                    & \textbf{Input type}          & \textbf{Words} &  \textbf{Precision}  \\ \midrule
Wang et al. \cite{wang2020dnrti}                                    & Sentence                         & 20                           &  0.89                \\ 
Wang et al. \cite{wang2022aptner}                                   & Sentence                         & 18                           &  0.83                \\ 
Fieblinger et al. \cite{fieblinger2024actionable}                   & Sentence                         & 54                           &  miss.                \\ 
Hu et al. \cite{hu2024llm}                                          & Paragraph                           & 106                          &  0.88                \\ 
\emph{The abstract of this paper}                                   & Paragraph                        & 174                          &  -                   \\
\emph{Emergency Directive} \\ 
\emph{21-01 (SolarWinds)} \cite{cisa_solarwinds_2021}    & Report                           & 1764                         &  -                  \\ \midrule
{\textbf{Our evaluation}} & Report & { \textbf{3009}} & {{\textbf{0.76}}} \\ \bottomrule
\end{tabular}
}
\end{center}
\end{table}

\smallbreak
\noindent\emph{RQ1: How do LLMs perform extraction on real CTI reports?} 

We use the open access datasets of reports by Di Tizio et al.~\cite{di2022software} which includes 350 reports on all Advanced Persistent Threats (APT)s reported by MITRE up to 2020 (and therefore should be known to the LLMs). Real, raw reports are two orders of magnitude larger than what is tested on the considered papers. The last column of Table \ref{tab:datasets_length} leaks the answer to our evaluation: LLMs are not great. Few-shot learning, prompt engineering, and fine-tuning do not significantly improve the results.

A different dimension of analysis is the consistency in the presence of repeated questions which is critical if the LLM is to be used as a chatbot. LLMs' output generation is potentially not deterministic \cite{lin2023generating,song2024good} and the possibility of receiving different results when extracting information from the same CTI report, poses severe risks in CTI such as for patch management. 

\smallbreak
\noindent\emph{RQ2: How to evaluate the CTI consistency of LLMs?}

Finally, even if the LLM may allucinate overconfidently \cite{yang2024can} we would like to have an estimate of its uncertainty which is normally required when reporting risks \cite{guikema2020artificial}. This analysis is also not reported in related work \cite{fieblinger2024actionable,hu2024llm,Kucsván2024330,song2024good}.

\smallbreak
\noindent\emph{RQ3: Are LLMs over(under)confident when making predictions in CTI?}

To address these challenges we: 1) design and deploy a novel evaluation pipeline to test the effectiveness, consistency and confidence calibration of LLMs for pre-attack CTI practices by extracting information from CTI reports and generating information regarding APTs, 2) run a validation experiment leveraging an existing open-source dataset consisting of 350 threat intelligence reports structured in STIX standard \cite{di2022software} as ground truth where we evaluate OpenAI, Google, and Mistral LLMs, 3) presents the result of the analysis on (i) the ineffectiveness of few-shot learning and fine-tuning for CTI, (ii) the inconsistency of LLMs, when used for information generation, and (iii) the low LLM confidence calibration in the extraction and generation of CTI information.

\subsection{Threat Model and Non-Goals}
A current focus of LLMs security research \cite{shayegani2023survey} is to implement adversarial attacks on LLMs such as prompt injection \cite{greshake2023not} and data poisoning \cite{chen2023janus} whose goal is to alter LLMs' behaviour by deceiving them into making incorrect predictions. The corresponding mitigation measures to prevent adversarial attacks \cite{chowdhury2024breaking}, include adding guardrails that can control LLMs output avoiding harmful outputs \cite{rebedea2023nemo}. 

While interesting, we consider it a \emph{non-goal} of our study, because there are already big security problems without calling attackers into play. We show that (uncompromised) LLMs may jeopardize the security of organizations using them to summarize CTI due to lack of consistency \cite{yang2024can} and calibration \cite{song2024good}. The intuitive cause is that real reports contain additional information besides the entities which the LLM must retrieve. While a report concerns a particular attack scenario which involved a specific APT, the vulnerabilities they exploited, and the attack vectors they used in that particular scenario, it often contains information about other APTs that might have used the same vectors in other occasions or other attack vectors used by the same APT in other scenarios. The irrelevant information in the report is easily confused as relevant and thus raises the number of False Positives (FPs) and False Negatives (FN).

\section{Examples of Unreliable LLMs in CTI}
\begin{figure}[h]
    \centering
    \includegraphics[width=0.90\textwidth]{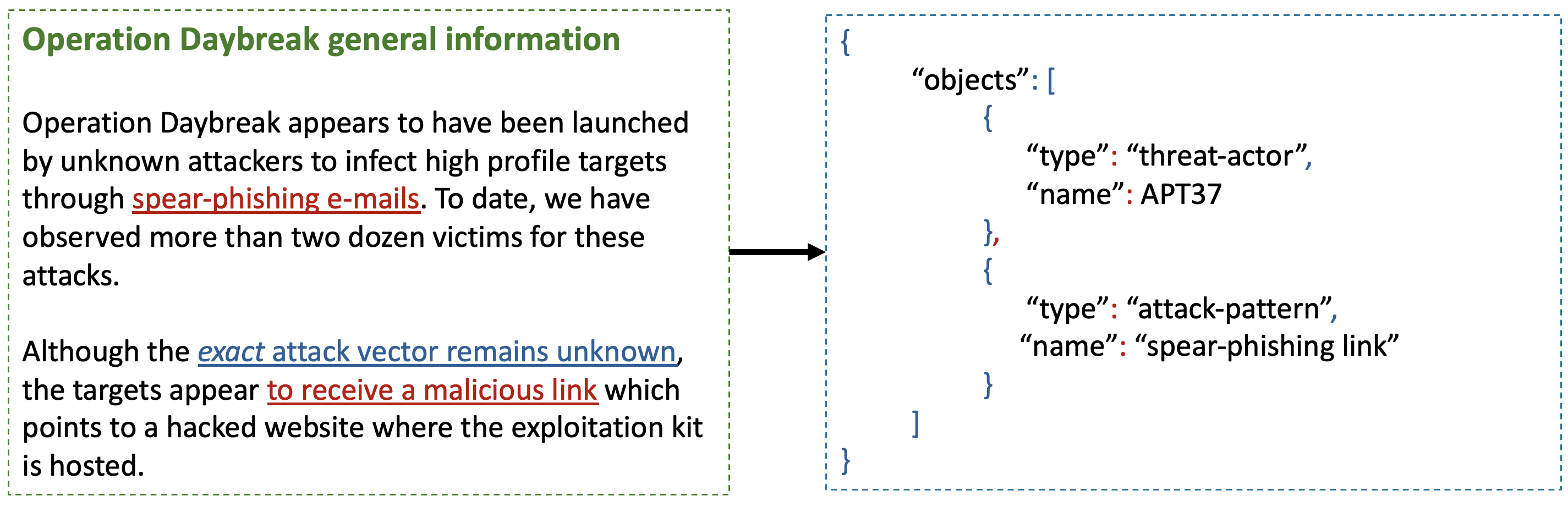}
    \caption{Excerpt of a CTI report (top) and STIX format (bottom). CTI reports can deceive tools by conveying contradictory information: \textbf{spear-phishing link} or \textbf{unknown attack vector}? The text fragment shown in the picture is extracted from a \href{https://securelist.com/operation-daybreak/75100/}{report} about a campaign by APT37.}
    \label{fig:contradictory_information}
\end{figure}

\begin{figure}
    \centering
    \includegraphics[width=0.90\linewidth]{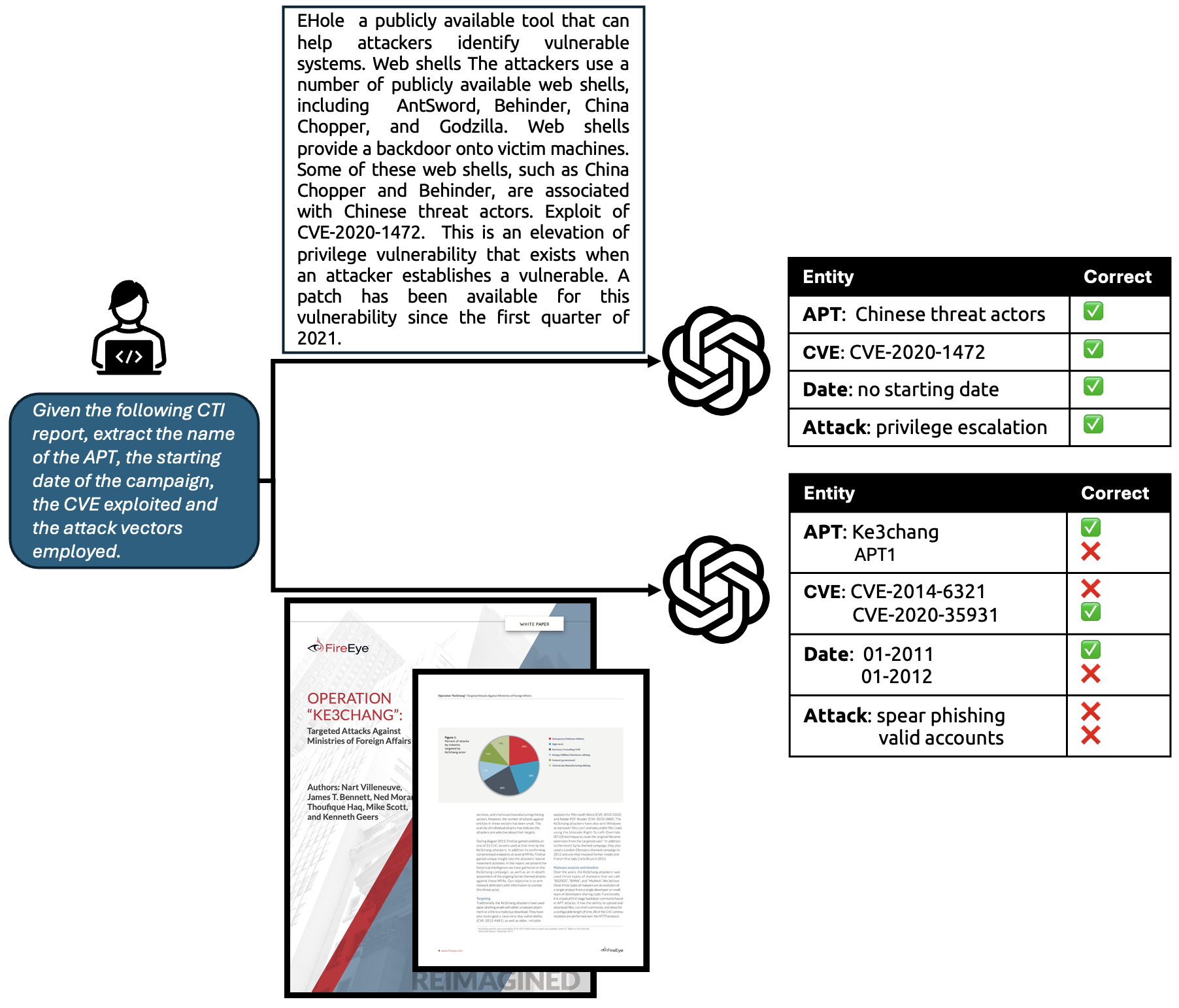}
    \begin{minipage}{\linewidth}{\footnotesize First, the analyst inputs the LLM with a paragraph of a threat report and the LLM extracts all the information contained in it. Then, the input consists of the entire report. Even though the input is better in terms of completeness, the result is worse as the number of FP and FN rises. The LLM is thus unable to distinguish the important information from the irrelevant ones.}
    \end{minipage}
    \caption{More Information, worse output.}
    \label{fig:scenario_rq1}
\end{figure}

\paragraph{\textbf{Longer reports, worse output.}}
The analyst prompts the LLM with the following instructions: \emph{Given the following CTI report, extract the name of the APT, the starting date of the campaign, the CVE of the vulnerabilities exploited and the attack vector employed.}
However, the final result can be compromised by the ambiguity of natural language CTI reports, shown in Figure \ref{fig:contradictory_information}, that makes automated extraction of CTI challenging. The correct (human) interpretation of this information is that spear-phishing links are the attack vector (as shown in the STIX format), but if taken out of context, the CTI report can be contradictory. This ambiguity, which can be reduced on small-size reports, forces to evaluate LLMs on real-size threat reports, to avoid the capacity of the models being overestimated due to the evaluation on small reports that do not represent the complexity of real CTI documents. 
Figure \ref{fig:scenario_rq1} shows the effects of report length on the LLM performance. The user prompts the LLM first by giving in input a paragraph of a threat report. The result reported by the LLM is perfect, as all the requested entities are extracted. The user then asks to repeat the same task with a complete report, and the result is disappointing, as the APT and starting date of the attack campaign should be one, thus raising the FP. 

\paragraph{\textbf{Ask twice and patch two different CVEs.}} 
It is possible that when the user prompts the LLM twice with the same instructions the information extracted differs between the two iterations \cite{song2024good,xiong2024efficient}, creating uncertainty in the patching steps. If we focus on the information regarding the vulnerabilities, this lack of determinism brings uncertainty concerning the CVE that is necessary to patch, thus delaying the fixing process. However, the delay in patching the right vulnerabilities will result in a higher chance of being attacked \cite{dissanayake2022and}. A key information is assessing the type of threat actor (e.g. nation-state actor or criminal organization), as that implies different attacker resources, and requires proportional defences \cite{mavroeidis2021threat}. Lack of consistency in this type of information can have ominous consequences.

\paragraph{\textbf{(Un)sure about the (right)wrong CTI.}}
When an LLM is employed it associates a number corresponding to the confidence about the predicted token(s). For example, the LLM may associate confidence of 0.90 for the \emph{APT} K3chang and 0.20 to the \emph{CVE} CVE-2014-6321. Model confidence is the parameter used to decide whether to accept or not the prediction when no dataset is available. Model confidence can only be trusted if the model is calibrated: the estimated probabilities are then representative of the true correctness likelihood \cite{guo2017calibration}.
When automated CTI pipelines blindly rely on the model confidence level, we do not know whether a high/low confidence value is justified or the model is over or underconfident. 
The model could return medium to high confidence levels about a specific APT extracted from a CTI report. An automated CTI pipeline does not know if the model is overconfident, and thus if the high confidence expressed reflects the true correctness likelihood. The model might be attributing an attack campaign from a report to a threat actor that probably did not implement it, \textit{introducing false positives}. 
The opposite case is also possible. If the model returns low confidence regarding an extracted vulnerability, an automated CTI pipeline will likely discard the prediction. If the model is underconfident, the confidence level does not reflect the true correctness likelihood. Thus, an automated pipeline may discard the prediction regarding a vulnerability that could be correct with a higher chance, \emph{raising the number of false negatives} and thus leading to failing to repair a vulnerability that can be attacked in the future \cite{ami2024false}. 

\section{Related Work}
\label{sec:related:work}
We searched on Scopus for all papers with CTI keywords and either NLP or LLM over the last five years.

\textbf{NLP and LLMs in CTI.} The increasing scale and complexity of cyber attacks lead to the necessity to automate CTI practices and sharing \cite{bairwa2023enhancing,bromander2021investigating,leite2023automated}. Thus, researchers invented formats to structure datasets containing CTI, which can be used to test new emerging methodologies for the automation of CTI analysis. Some datasets can be employed for specific tasks such as named-entity recognition (NER) in CTI, such as \cite{liu2023apttoolner} and \cite{wang2022aptner}. Other general-purpose datasets can be employed for KG extraction \cite{alam2023looking,wang2020dnrti}. Husari et al. \cite{husari2019learning}, propose an NLP-based analysis to reconstruct the chains of the attacks composed by the actions performed by the APT. In contrast, Zhang et al. \cite{zhang2021ex} propose EX-Action a framework to extract threat actions from cyber threat reports. Abdi et al. \cite{abdi2023automatically} propose an NLP-based system to automatically highlight a CTI report with the responsible actor. Important tools offered by the world of NLP to researchers consist of PLMs and LLMs. One of the first activities to which PLMs were applied was called entity recognition in cyber threat reports. Quiao et al. \cite{qiao2023improving} employs BERT-based models to annotate nine different categories of entities. Researchers also employed LLMs to help analysts respond to incoming attacks or threat campaigns, for instance automating the labelling of network intrusion detection systems rules, and the consultation of frameworks such as ATT\&CK and D3FEND \cite{daniel2024labeling}. Researchers also tested the capacity of LLMs to automate the generation of threat reports \cite{perrina2023agir} or to retrieve recovery steps from threat reports \cite{Kucsván2024330}. Finally, researchers employ PLMs \cite{jo2022vulcan,marchiori2023stixnet,sewak2023crush} and LLMs \cite{liu2023constructing} to extract entities and connections from threat reports, thus automating the operation of KG extraction. Fieblinger et al. \cite{fieblinger2024actionable}, Hu et al. \cite{hu2024llm}, and \cite{liu2023constructing} tests LLMs on the task of KGs from cyber threat reports reporting promising performances. LLMs are also starting to be employed as CTI assistants to assign the correct CWE to a specific CVE \cite{ferrag2024generative,madden2024genai}. Current approaches that employ LLMs for CTI extraction from threat reports show promising performance. However, they test LLMs on single sentences and report paragraphs that are less complex than real-size reports. Moreover, even though LLMs are starting to be used as CTI assistants, there is still the necessity to perform proper evaluation. In this work, we address this gap by evaluating LLMs in the task of information extraction on real-size reports and in the task of CTI assistants that will build the APT profile given the given APTs' names. 

\textbf{Consistency quantification in LLMs.} Considering the lack of determinism of LLMs \cite{huang2023survey,song2024good}, a complete branch of research is dedicated to quantifying the consistency characterising their outputs \cite{fadeeva2023lm,xiong2024efficient,ye2024benchmarking}. It is possible to highlight two categories of consistency quantification in the realm of LLMs, where the first concerns the general characteristics of LLMs as free-form generation, while the second looks a the capacities of the LLMs regarding more limited and close-ended applications such as classification and information extraction. Examples of the first case are the scientific research published by Kuhn et al. \cite{kuhn2023semantic} who introduce the concept of semantic entropy, helping to measure the uncertainty regarding sentences and outputs which have the same meaning, and Lin et al. \cite{lin2023generating} who propose different metrics to evaluate the uncertainty in black-box LLMs. Examples of the second path of research can be seen in the scientific articles written by Wang et al. \cite{wang2022uncertainty}, concerned with uncertainty quantification in the realm of text-regression, and in the research carried on by Jiang et al. \cite{jiang2021can} and Kamath et al. \cite{kamath2020selective}, concerning uncertainty quantification in the case of PLMs employed for question answering. Current approaches in CTI measure performance with point value metrics and do not consider issues due to a lack of determinism. We address this gap and present an evaluation step that quantifies LLM consistency.

\begin{figure}[t]
     \centering
     \begin{subfigure}[b]{0.45\textwidth}
         \centering
         \includegraphics[width=\textwidth]{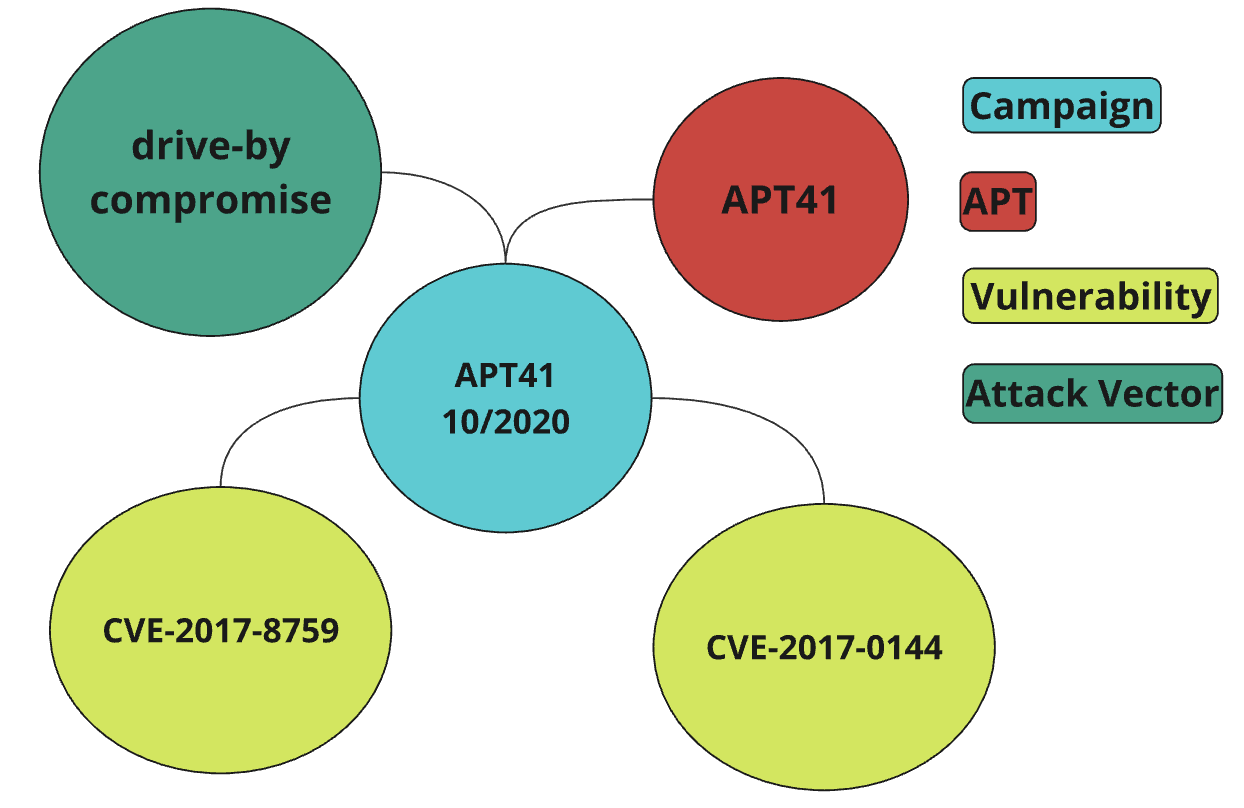}
         \caption{KG extracted from CTI reports.}
         \label{fig:campaign_graph}
     \end{subfigure}
     \hfill
     \begin{subfigure}[b]{0.45\textwidth}
         \centering
         \includegraphics[width=\textwidth]{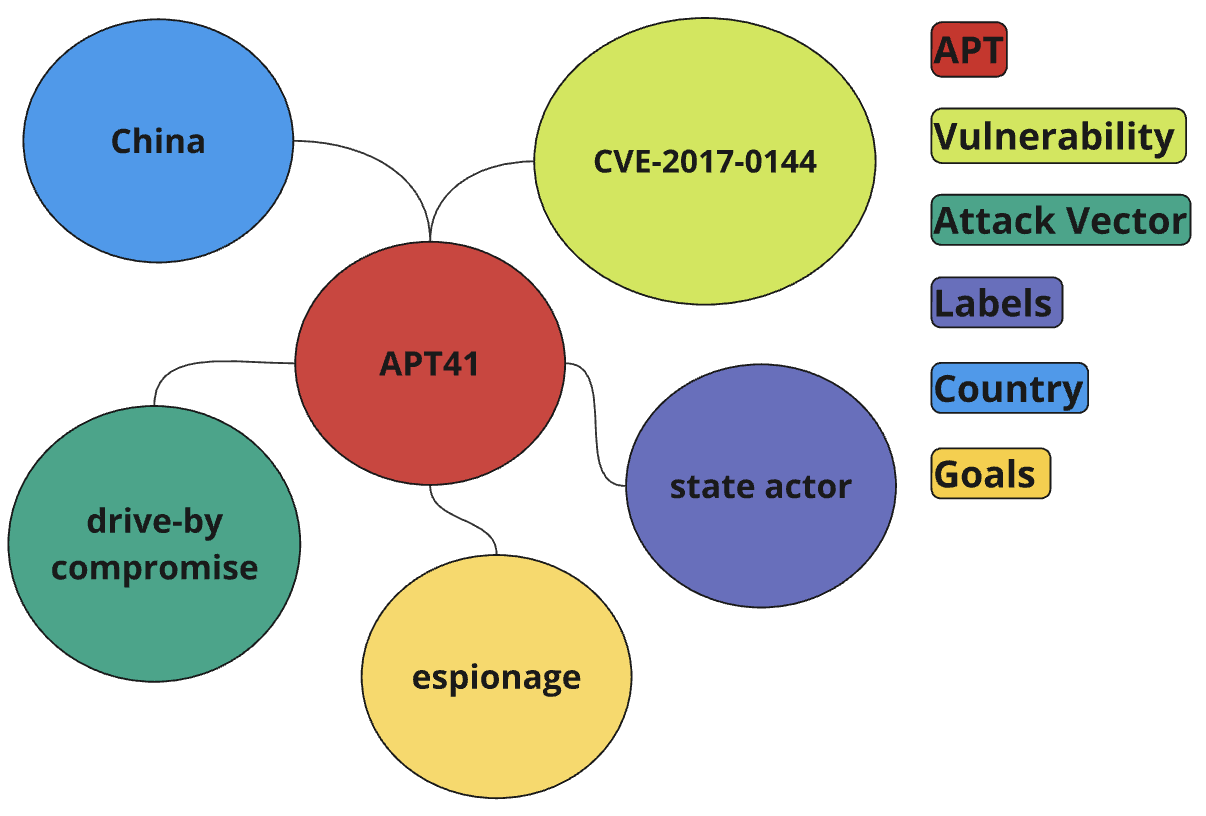}
         \caption{KG from information generation.}
         \label{fig:apt_graph}
     \end{subfigure}

     \begin{minipage}{\linewidth}\footnotesize
         Figure \ref{fig:campaign_graph} and Figure \ref{fig:apt_graph} picture the graphs created when extracting information from cyber threat reports and generation information from the names of APTs. The graph resulting from information extraction is characterized by the entity \emph{Campaign} that is composed by the name of the APT and by the starting date of the campaign, while the graph resulting from information generation is characterized by the \emph{country} of origin of the APT, by its \emph{goals} and by its \emph{labels} which correspond to the type of APT. 
     \end{minipage}
     \caption{KG extracted (a) and generated (b) when performing CTI tasks.}
\end{figure}

\textbf{Calibration in LLMs.}
Lack of calibration is a characteristic of contemporary Deep Learning (DL) models \cite{guo2017calibration}. However, even though new methods to improve calibration for LLMs have been suggested \cite{yang2024can}, in security this is still an unexplored problem. To the best of our knowledge, only one article explored LLMs' calibration and how to improve it for code generation \cite{spiess2024calibration}. Lack of calibration can have consequences in case LLMs are deployed in the absence of a labelled dataset, where the prediction confidence is used to choose whether to consider or discard their predictions. No previous work analysed calibration when LLMs are involved in CTI tasks. We address this gap by performing a calibration analysis to determine whether LLMs are (under)overconfident.

\section{Overview of the Approach} 
We design our five-step evaluation pipeline for two critical tasks.

\textbf{Information extraction.} When the LLM is given a \emph{single} unstructured CTI report it is asked to extract entities from this report such as the APT the report is about or the attack vector it has exploited according to the report. This task is essential to transform natural language text into a structured format such as STIX. Figure \ref{fig:campaign_graph}, shows an example of the expected outcome. 

\begin{figure*}[t]
    \centering
    \includegraphics[width=\textwidth]{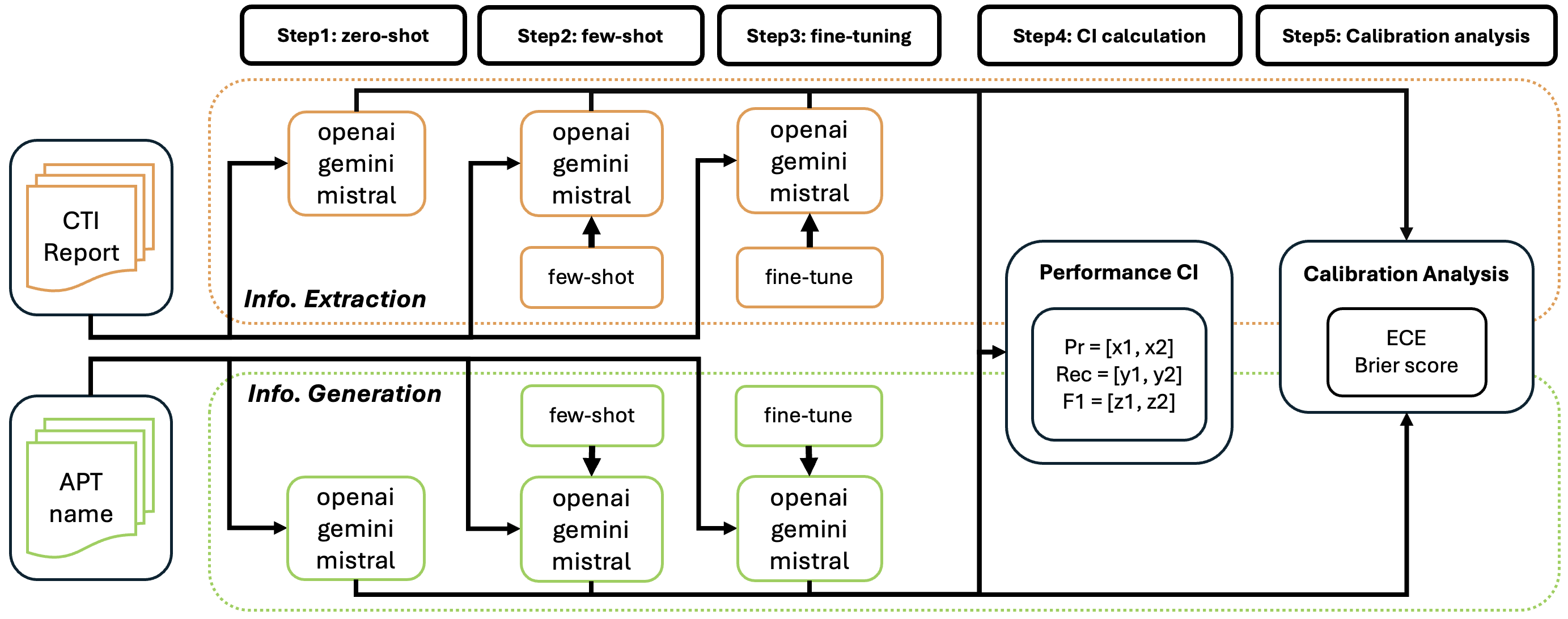}
    \begin{minipage}{\linewidth}{\footnotesize
    Each evaluation step involves two selected CTI tasks: information extraction from threat reports and information generation from APT names, respectively represented by the orange and the green frame. The \nth{1} step evaluates LLMs with zero-shot learning, the \nth{2} with few-shot learning, and the \nth{3} after fine-tuning the model. Performance is registered as point values. In the \nth{4} step, we quantify LLM performance consistency by calculating performance confidence intervals (CI). During the \nth{5} step, we compute the expected calibration error (ECE) and Brier Score (BS).}
    \end{minipage}
    \caption{Illustration of the LLMs evaluation.}
    \label{fig:pipeline}
\end{figure*} 

\textbf{Information generation.} When the LLM is asked to provide information related to an APT (e.g., the used attack vectors, the country of origin etc.) based on the knowledge embedded in its weights. Figure \ref{fig:apt_graph} shows an example output.

Figure \ref{fig:pipeline} shows the evaluation steps. The first step consists of evaluating each LLM with zero-shot learning. The second step uses few-shot learning. To control for \emph{prompt overfitting} \footnote{With prompt overfitting, we refer to the case in which an LLM can perform well on the data from which the few-shot examples that enrich the prompt are gathered, but it is not able to use those few-shot examples to generalize its performance to data samples that do not contain those same few-shot examples \cite{cho2024make}.}, we evaluate the LLMs on the part of the dataset from which we do not gather any few-shot examples, mimicking the division between train and test datasets, and checking whether the model can generalize from given examples. The same approach is performed for fine-tuning: we fine-tune the LLM on the dedicated dataset section and then test it on the remaining part of the dataset. The fourth step calculates the confidence intervals (CIs) regarding the performance, and quantifies the oscillation in model performance, an aspect overlooked in previous work evaluating LLMs on CTI tasks \cite{fieblinger2024actionable,hu2024llm}.

Precision and recall are sufficient only when models can be evaluated with a labelled dataset. When models must be used in real-world scenarios, knowledge regarding calibration is needed to know whether the predictions can be trusted  \cite{spiess2024calibration}. The fifth step consists of deriving model confidence by analyzing the log probabilities generated by the LLMs. We can thus check whether the models are calibrated and whether few-shot learning and fine-tuning improve the calibration level, an aspect overlooked in previous work on LLMs for CTI \cite{fieblinger2024actionable,hu2024llm,Kucsván2024330,Tihanyi2024296}.

\section{Evaluation Metrics}

\textbf{Traditional Metrics of Performance.} For evaluating the impact of the size of the report we use the traditional metrics used in the literature (See Section~\ref{sec:related:work} and Table~\ref{tab:datasets_length}): \emph{Precision} $P=\frac{TP}{TP+FP}$ is the portion of extracted elements of a specific class which were correctly extracted; \emph{Recall} $R=\frac{TP}{TP+FN}$ is the portion of elements of a specific class which have been extracted by the model; \emph{F1} is the harmonic mean of precision and recall.

\textbf{Confidence intervals.} We measure the consistency of LLM output when prompted several times with identical inputs. To obtain measures beyond point estimates, we build confidence intervals from our observations by relying on a multi-sample method \cite{kuhn2023semantic}. We draw sample values from the population with bootstrapping \cite{efron1992bootstrap} with replacement, and calculate the sample mean. We repeat this process where $n$ is the population size and $k$ is the size of the sample drawn. Then, we create a list with the sample means, the \nth{5} and \nth{95} percentile (lower and upper bound of the interval).

\textbf{Calibration metrics.} We evaluate the LLMs calibration with expected calibration error (ECE) and Brier score (BS). The ECE \cite{naeini2015obtaining} and BS \cite{brier1950verification}, are two measures of calibration that quantify the deviation from perfect calibration. From Jiang et al. \cite{jiang2021can}, given an input $X$ and true output $Y$, a model output $\hat{Y}$, and a probability $PN(\hat{Y}|X)$ calculated over this output, a perfectly calibrated model satisfies the condition:

\begin{equation}
    P(\hat{Y}=Y | PN(\hat{Y}|X)=p)=p, \forall p\in[0,1]
\end{equation}

which is that the confidence \emph{p} of the model, corresponding to the calculated probability of its prediction that the output is $\hat{Y}$, equals the empirical fraction \emph{p} of the cases where the actual output $Y$ correctly matches the prediction $\hat{Y}$ \cite{spiess2024calibration}. The best obtainable value for ECE and BS is zero, indicating perfect calibration. The higher the value is, the lower the calibration.  

\section{Experimental Pipeline}

\begin{table*}
\def\makespace{\rule{0.pt}{3ex}}
\begin{center}
\caption{Prompt engineering techniques employed taken from \cite{chen2023unleashing,OpenAIdocPrompt}.}
\label{tab:prompt_enegineer_practices}\footnotesize
\begin{tabular}{p{0.15\linewidth}@{~~}p{0.25\linewidth}@{~~}p{0.6\linewidth}}
\toprule
{\textbf{Technique}} & \textbf{Description} &\textbf{Example} \\ \midrule
{Role specification} & Instructing the LLM on the role. & \texttt{You are a Cyber Threat Intelligence (CTI) analyst.}\\ \makespace
Step specification & Specify the steps required to accomplish a task. & \begin{minipage}[t]{0.3\linewidth}\begin{verbatim}
Step 1 - Extract the starting date of 
the campaign, the Advanced Persistent 
Threat (APT), the CVE codes of the 
vulnerabilities exploited by the APT ...
\end{verbatim}
\end{minipage} \\\makespace 
 Input subdivision & Split the steps into different and separated sections. &
\begin{minipage}[t]{0.3\linewidth}
\begin{verbatim}
Step 2 - Return the information filling in 
this JSON format: 
"nodes": {
    "APT": [{"name": ""}], 
    "attack_vector": [{"name": ""}], ...
\end{verbatim}
\end{minipage}
\\ \makespace
{World closing} & Reducing the values an LLM can assign to an entity, indicating the possible alternatives. & \texttt{The name of the attack vector can only be one of the following: drive-by compromise, supply chain compromise, spear-phishing via service, spear-phishing attachment ...} \\ \makespace
{Few-shot learning} & Providing a small number of labelled examples to the LLM from which it can generalize. & \begin{minipage}[t]{0.3\linewidth}
\begin{verbatim}
Examples to understand which attack vector 
the APT used. 
- ... employed legitimate user credentials 
      to access its targets: valid accounts.
- ... has been linked to a watering hole 
      attack: drive-by compromise. ... 
\end{verbatim}
\end{minipage} \\ \bottomrule
\end{tabular}
\end{center}
\end{table*}


\textbf{Prompt preparation.} To ensure the quality of the prompts we rely on the practices from prompt engineering (Table \ref{tab:prompt_enegineer_practices}). The techniques used are role specification, which instructs the LLM to embody the role of CTI analyst, input subdivision and step specification, which allows to specify a different task for each section of the prompt. We apply world closing by reducing, based on the dataset in \cite{di2022software}, the possible values that entities can be assigned to by the LLM during information extraction and generation. 

\textbf{Zero-shot learning.}
We evaluate the capacity of the LLMs to perform the selected tasks without providing any examples through the prompt to help the LLM to extract and classify the information contained in the threat reports or the description of the APTs \cite{kojima2022large}. 
To perform information extraction, the LLM receives in input the prompt which instructs it regarding the entities to extract, and the cyber threat report from which they are to be extracted. For information generation, the LLM receives in input the name of the APT, the description of the APT, and the instructions indicating the information to recover. 

\textbf{Few-shot learning and fine-tuning.} In the second and third steps, we repeat the tasks, by applying few-shot learning \cite{brown2020language} and fine-tuning \cite{mosbach2023few}, measuring to what extent these techniques affect the performance of the LLMs for information extraction and generation. To implement few-shot learning for information extraction and generation, we respectively extract few-shot examples from threat reports and APTs' descriptions. We evaluate the LLMs on the dataset section from which few-shot examples were not extracted, to ensure that prompt overfitting is avoided. To fine-tune LLMs we randomly select a portion of the threat reports and APTs' descriptions and create the training dataset composed of the prompts and the correct answers to be generated. At the end of the fine-tuning, we evaluate the models on the test dataset, consisting of the dataset section not used for the training. We rely on supervised fine-tuning (SFT) \cite{pareja2024unveiling} instead of more recent techniques based on reinforcement learning, such as Reasoning with REinforced Fine-Tuning (REFT) \cite{luong2024reft} as they suffer from reward hacking.

\textbf{Generation of confidence intervals (CI).}
Since LLMs suffer from a lack of determinism \cite{song2024good}, our evaluation measures their capacity to generate consistent results over multiple iterations. To this aim, we rely on a multi-sample approach \cite{kuhn2023semantic,xiong2024efficient}, by repeating the process of information extraction and information generation multiple times on the same input, registering the performance and then generating CIs. For each task, we re-prompt each LLM ten times on the same input, with \emph{temperature=0} and the same seed to guarantee maximum determinism, derive precision, recall, and f1-score, and then employ the bootstrapping method \cite{efron1992bootstrap} to empirically build CIs. The bootstrapping method avoids assumptions over the distribution of data samples represented by metrics values gathered during the iterations. We re-prompt the LLM ten times to balance the necessity of checking for lack of consistency in model output with expense limits that would allow researchers to reproduce experiments. 

\textbf{LLM calibration analysis.} In the last step, we analyse whether LLMs are calibrated regarding the selected CTI tasks. We perform this analysis for zero-shot learning, few-shot learning and the fine-tuned model, to check whether few-shot learning and fine-tuning help in improving model calibration in CTI tasks \cite{kapoor2024calibration,spiess2024calibration}. We implement calibration analysis by extracting the log probabilities that the LLM assigns to the tokens composing the generated response. Considering that the response is in the JSON format, and each field of the JSON file corresponds to a different entity, we isolate the tokens and the probabilities of each section of the JSON file corresponding to a specific entity, multiply the token probabilities of that JSON file section, and derive the overall confidence for each entity. Finally, we calculate the ECE \cite{naeini2015obtaining} and BS \cite{brier1950verification}, to measure the deviation from the ideal confidence level.

\section{Dataset and Models}
\label{sec:dataset}

\textbf{Dataset selection.}
For our evaluation, we use the APT dataset from Di Tizio et al. \cite{di2022software}. We report in Table \ref{tab:main_dataset_statistics} the characteristics of interest for our study. 

\begin{table}[t]
\begin{center}
\caption{Dataset \cite{di2022software} key Indicators.}
\label{tab:main_dataset_statistics}
\resizebox{0.70\columnwidth}{!}{%
\begin{tabular}{cc}
\begin{tabular}{l@{}r}
\toprule
\textbf{Entities} & \textbf{Num} \\\midrule
{\# of reports}    & 350  \\
{\# Campaign}       & 350  \\
{\# APT}            & 86   \\
{\# Vulnerability}  & 123  \\
{\# Attack Vector}  & 170  \\
{\# Country}        & 17   \\
{Max vulns/campaign} & 6 \\
{Max attack vectors/campaign} & 4 \\ \bottomrule
\end{tabular}
&
\begin{tabular}[b]{lrr}
\toprule
& \multicolumn{2}{c}{\textbf{Report Size}} \\
            & \textbf{Mean} & \textbf{Max} \\ 
            \midrule
{\# words}  & 3\,009 & 21\,569 \\
{\# tokens} & 4\,002 & 27\,794 \\ \midrule
\end{tabular}
\end{tabular}
}
\end{center}
\vspace*{-\baselineskip}
\end{table}

We have selected the dataset as it is open source, it has been manually curated for correctness and presents characteristics in terms of length of reports and heterogeneity that allow us to test the LLMs on real-size CTI reports. The dataset guarantees heterogeneity by selecting 86 of the 163 APT groups on the MITRE ATT\&CK. The selected APTs are the ones that launched at least one campaign during 2008 and 2020. The amount of attack campaigns and thus of CTI reports is 350. 
Each CTI report is associated with the following entities that can be extracted:
\emph{APT}, \emph{Campaign}, \emph{Vulnerability}, and \emph{Attack vector}. The \emph{APT} is the actor responsible for the attacks, the \emph{Campaign} is composed of the name of the APT responsible for the attacks and its starting date, the \emph{Vulnerability} corresponds to the \emph{CVE} codes of the vulnerabilities exploited by the APT, and the \emph{Attack Vectors} are the techniques employed by the threat actor. \\ The origin and structure of the textual sources widely vary: from cyber threat reports written by cybersecurity providers to blog posts shared by cybersecurity enthusiasts. Other than the threat reports, the dataset is equipped with information regarding the 86 APTs. Each APT is associated with the \emph{country} of origin of the APT, the \emph{label} of the APT which refers to whether the APT is a criminal, a nation-state actor or a spy, its \emph{goals} such as espionage, the \emph{vulnerabilities} they exploited over their campaigns, and the \emph{attack vectors} employed.

\textbf{Dataset division.}
We randomly select 70\% of the dataset, thus mimicking the splitting which is typical of ML and DL model validations. In ML and DL, the splitting would have been in training, validation, and testing respectively in portions of 70\%, 20\%, and 10\%. Since we do not train from scratch our model we assign 70\% to the few-shot examples section and fine-tuning and 30\% for the testing. We employ this division to avoid overfitting both for fine-tuning and for few-shot learning, if the few-shot examples are gathered directly from the dataset on which testing is conducted \cite{cho2024make}. 

\textbf{Models.} We implement all the experiments by employing LLMs which are state-of-the-art at the time of writing, that can be fine-tuned, and for which it is available the JSON modality: \textit{gpt4o} from OpenAI, \textit{gemini-1.5-pro-latest} from Google, and \emph{mistral-large-2} from Mistral, that can digest even the longest threat report given their large context windows (128k, 2M, and 128k tokens).

\section{RQ1. Performance of LLMs in CTI}

Table \ref{tab:table_extraction_rq1} and Table \ref{tab:table_generation_rq1} (in the Appendix) show the results related to the first research question for entity extraction and entity generation respectively. We test LLMs with zero-shot learning, few-shot learning, and after fine-tuning.

\textbf{Information extraction.} Focusing on information extraction, Table \ref{tab:table_extraction_rq1} shows that zero-shot learning does not bring positive performance. Focusing on the recall in the best case is equal to 0.90, when \textit{gemini} and \textit{mistral} retrieve CVEs, meaning that 10\% of the vulnerabilities will be overlooked. Recall can be as low as 0.72 when \textit{gpt4o} is employed to retrieve \emph{campaigns} entities, thus overlooking 28\% of the campaigns. Even more worrying is the lack of benefits of few-shot learning and fine-tuning. When applying few-shot learning the performance can decrease below the performance obtained with zero-shot learning. The maximum decrease is obtained when considering the \emph{APT} specifically for the models \textit{gemini} and \textit{mistral} for which the performance decreases by 7.87\% from 0.89 to 0.82. The decrease in performance means that employing LLMs for CTI tasks will cause overlooking \emph{APT} names, \emph{CVEs}, and \emph{campaign} and thus to an error in comprehension of the scenario and in the attack attribution. The same and even worsened negative trend is evident when analyzing the performance model after fine-tuning. The table shows that the recall decreases as low as 0.58 from 0.72 when \emph{gpt4o} is applied to retrieve \emph{campaign}, leading to overlooking the 42\% of the \emph{campaign} entities. However, the worst decrease is obtained when \textit{gpt4o} and \textit{mistral} are used to extract \emph{APT} names with precision and recall decreasing from 0.87 to 0.68, with a decrease of 21.84\%. 

\begin{table*}[t]
\caption{Results for information extraction.}
\label{tab:table_extraction_rq1}
\begin{center}
\resizebox{0.80\columnwidth}{!}{%
\begin{tabular}{llrrrrrrrrr}
\toprule
    && 
    \multicolumn{3}{c}{\textbf{zero-shot}} & 
    \multicolumn{3}{c}{\textbf{few-shot}} & 
    \multicolumn{3}{c}{\textbf{fine-tuning}} \\
    &  \textbf{Model} & 
    \textbf{P} & 
    \textbf{R} & 
   {\textbf{F1}} & 
    \textbf{P} & 
    \textbf{R} & 
   {\textbf{F1}} & 
    \textbf{P} & 
    \textbf{R} &
    \textbf{F1} \\ \midrule
campaign
    & 
gpt4o & 
    0.72 & 
    0.72 & 
   {0.72} & 
    0.72 & 
    0.72 & 
   {0.72} &
    0.58 & 
    0.58 & 
    0.58 \\
& gemini & 
    0.77 & 
    0.77 & 
   {0.77} & 
    0.73 & 
    0.73 & 
   {0.73} & 
    0.61 & 
    0.61 & 
    0.61 \\
& mistral & 
    0.74 & 
    0.74 & 
   {0.74} & 
    0.69 & 
    0.69 & 
   {0.69} & 
    0.58 & 
    0.58 & 
    0.58 \\ \cmidrule{2-11}
APT &
gpt4o & 
    0.87 & 
    0.87 & 
   {0.87} & 
    0.84 & 
    0.84 & 
   {0.84} & 
    0.68 & 
    0.68 & 
    0.68 \\
& gemini & 
    0.89 & 
    0.89 & 
   {0.89} &
    0.82 & 
    0.82 & 
   {0.82} &
    0.80 & 
    0.80 & 
    0.80 \\
& mistral & 
    0.89 & 
    0.89 & 
   {0.89} & 
    0.82 & 
    0.82 & 
   {0.82} & 
    0.68 & 
    0.68 &
    0.68 \\ \cmidrule{2-11}
CVE & 
gpt4o & 
    0.67 & 
    0.87 & 
   {0.76} & 
    0.74 & 
    0.92 & 
   {0.82} & 
    0.71 & 
    0.69 & 
    0.70 \\
&gemini & 
    0.69 & 
    0.90 & 
   {0.78} &  
    0.75 & 
    0.89 & 
   {0.81} & 
    0.81 & 
    0.63 & 
    0.71 \\ 
& mistral & 
    0.72 & 
    0.90 & 
   {0.80} &
    0.79 & 
    0.91 & 
   {0.85} &
    0.71 & 
    0.69 & 
    0.70 \\ \cmidrule{2-11}
attack vector &    
gpt4o & 
    0.53 & 
    0.75 & 
   {0.62} &
    0.44 & 
    0.77 & 
   {0.56} &
    0.69 & 
    0.65 & 
    0.67 \\
& gemini & 
    0.68 & 
    0.74 & 
   {0.71} & 
    0.71 & 
    0.78 & 
   {0.74} & 
    0.89 & 
    0.84 & 
    0.87 \\
& mistral & 
    0.67 & 
    0.83 & 
   {0.74} &
    0.67 & 
    0.85 & 
   {0.75} &
    0.69 & 
    0.65 & 
    0.67 \\ \bottomrule
\end{tabular}}
\end{center}
\end{table*}

\textbf{Information generation.} Table \ref{tab:table_generation_rq1} (in the Appendix) shows the results for information generation. Generally, the recall is considerably low, with the lowest level registered when the LLM is used to generate the type of \emph{APT}, with 0.02 reached by \textit{gemini} and \textit{mistral}. The effect of few-shot learning and fine-tuning is limited and can also be detrimental, with the only exception represented by the \emph{goals} entity. For few-shot learning, the maximum decrease in performance is obtained when \textit{mistral} is used to generate the country of the threat actor, starting with precision and recall of 0.78 and ending with precision and recall of 0.64, and thus 36\% of APTs' countries are wrongly predicted. Low performance is obtained when the LLM is fine-tuned. The worst case is represented by the application of \textit{gpt4o} and \textit{mistral} to generate \emph{CVEs}, when the registered recall is 0.00, meaning that after fine-tuning the LLM is completely unable to generate and highlight the CVEs exploited by threat actors.

\section{RQ2. Consistency of LLMs output}
Here we analyze the results concerning the consistency quantification for LLMs generated output. The larger the CI computed the lower the consistency. Thus, we attribute an LLM perfect determinism or perfect consistency when the width of the CI is zero, indicating that the LLM returns the same output given the same input over multiple iterations. 

Table \ref{tab:intervals_node_extraction} and Table \ref{tab:intervals_node_generation} (in the Appendix) show the performance CI for the two tasks. For both tasks and most entities the LLMs are not consistent. Between the two tasks, information generation shows a greater lack of consistency. For information extraction in the worst case, the difference between the lower and upper bound of the CI is $0.02$. The highest width can only be found between fine-tuned models, for instance when \textit{gemini} is used to retrieve \emph{campaign} with a precision and recall CI of [0.58, 0.60] with a difference between the lower and upper bound of 3.39\% or when \textit{mistral} is used to retrieve \emph{CVE} codes, with precision and recall of [0.72, 0.74] and thus a difference between lower and upper bound of 2.74\%. 

For the task of information generation, the maximum difference between the lower and upper bound is $0.06$, when few-shot learning is applied to \textit{gemini} to generate \emph{CVE} codes with a recall CI of [0.19, 0.25] and thus a 27.27\% percentage difference between the lower and upper bound. 

\begin{table*}[t]
\caption{Performance CI calculated with the bootstrapping techniques, for information extraction. LLMs lack complete determinism.}
\label{tab:intervals_node_extraction}
\begin{center}
\resizebox{0.90\columnwidth}{!}{%
\begin{tabular}{llcccccc}
\toprule
           && \multicolumn{3}{c}{\textbf{few-shot}}                                      & \multicolumn{3}{c}{\textbf{fine-tuning}}                  \\ 
           & \textbf{Models}           & \textbf{P}       & \textbf{R}       &{\textbf{F1}}        & \textbf{P}       & \textbf{R}       & \textbf{F1}         \\
\midrule
campaign
& gpt4o & {[}0.69, 0.70{]} & {[}0.69, 0.70{]} &{{[}0.69, 0.70{]}} & {[}0.55, 0.56{]} & {[}0.55, 0.56{]} & {[}0.55, 0.56{]} \\
& gemini     & {[}0.73, 0.74{]} & {[}0.73, 0.74{]} &{{[}0.73, 0.74{]}} & {[}0.58, 0.60{]} & {[}0.58, 0.60{]} & {[}0.58, 0.60{]} \\
& mistral    & {[}0.67, 0.68{]} & {[}0.67, 0.68{]} &{{[}0.67, 0.68{]}} & {[}0.55, 0.56{]} & {[}0.55, 0.56{]} & {[}0.55, 0.56{]} \\ \cmidrule{2-8}
APT
& gpt4o & {[}0.84, 0.84{]} & {[}0.84, 0.84{]} &{{[}0.84, 0.84{]}} & {[}0.66, 0.68{]} & {[}0.66, 0.68{]} & {[}0.66, 0.68{]} \\
& gemini     & {[}0.84, 0.84{]} & {[}0.84, 0.84{]} &{{[}0.84, 0.84{]}} & {[}0.78, 0.79{]} & {[}0.78, 0.79{]} & {[}0.78, 0.79{]} \\
& mistral    & {[}0.82, 0.82{]} & {[}0.82, 0.82{]} &{{[}0.82, 0.82{]}} & {[}0.66, 0.68{]} & {[}0.66, 0.68{]} & {[}0.66, 0.68{]} \\  \cmidrule{2-8}
CVE
& gpt4o & {[}0.74, 0.74{]} & {[}0.89, 0.90{]} &{{[}0.81, 0.82{]}} & {[}0.72, 0.74{]} & {[}0.71, 0.74{]} & {[}0.72, 0.74{]} \\
& gemini     & {[}0.74, 0.75{]} & {[}0.89, 0.89{]} &{{[}0.80, 0.81{]}} & {[}0.80, 0.81{]} & {[}0.62, 0.63{]} & {[}0.70, 0.71{]} \\
& mistral    & {[}0.80, 0.81{]} & {[}0.91, 0.91{]} &{{[}0.85, 0.85{]}} & {[}0.72, 0.74{]} & {[}0.72, 0.74{]} & {[}0.72, 0.74{]} \\ \cmidrule{2-8}
attack vector
& gpt4o & {[}0.42, 0.43{]} & {[}0.77, 0.77{]} &{{[}0.54, 0.55{]}} & {[}0.71, 0.73{]} & {[}0.67, 0.69{]} & {[}0.69, 0.71{]} \\
& gemini     & {[}0.72, 0.73{]} & {[}0.77, 0.78{]} &{{[}0.74, 0.75{]}} & {[}0.90, 0.91{]} & {[}0.85, 0.86{]} & {[}0.87, 0.88{]} \\
& mistral    & {[}0.66, 0.66{]} & {[}0.83, 0.84{]} &{{[}0.73, 0.74{]}} & {[}0.71, 0.73{]} & {[}0.71, 0.73{]} & {[}0.71, 0.73{]} \\ \bottomrule
\end{tabular}}
\end{center}
\end{table*}

\section{RQ3. Analysis of LLMs calibration level}
The calibration analysis is performed by reporting the values of ECE and BS, which indicate the deviation of the model from perfect calibration. The best possible value for ECE and BS is zero, thus the higher their value the lower the calibration. Table \ref{tab:calibration_data_extraction} shows the results related to the calibration analysis. 

\textbf{Information extraction.} Few-shot learning is detrimental to the model calibration, as seen by the variations in the ECE and BS. An increase in ECE and BS is seen when considering \emph{campaign} entity in which the ECE and BS increase from 0.25 and 0.26 to 0.26 and 0.28 respectively. The same can be seen for the \emph{attack vector} where ECE and BS increase from 0.13 and 0.46 to 0.19 and 0.49. The \emph{CVE} entity presents the worst increase for ECE and BS, as they increase from 0.28 and 0.32 when zero-shot learning is used to 0.35 and 0.37 when employing few-shot learning. 

Also fine-tuning worsens the LLM calibration. The only exception that shows improvement is the \emph{CVE} entity, which diminishes both the ECE and BS when evaluating the fine-tuned LLM. All the other entities worsen their performance, with the maximum increase reached by the \emph{campaign} entity, which registers an ECE and BS of 0.48 when calculated on the fine-tuned model performance. An increase of 92\% in ECE and 84.62\% for BS, compared to the ECE and BS registered when zero-shot learning is applied. 

\textbf{Information generation.} The LLM employed for information generation offers a similar scenario. The only exception is the \emph{goals} entity, in which few-shot learning and fine-tuning improve the ECE and BS. All other entities worsen their performance. An example of this pattern is the \emph{labels} entity, which registers an ECE and BS of 0.45 and 0.44 when zero-shot learning is used, which rises to 0.57 and 0.53 when few-shot learning is used. This pattern is evident also for \emph{attack vector} where the ECE raises from 0.47 to 0.48. 

More concerning are the effects of fine-tuning on model calibration. The worst case is highlighted by the \emph{CVE} and \emph{attack vector} entity. For CVE the fine-tuned LLM registers an ECE and BS of 0.91 and 0.98 signalling a complete misalignment between performance and confidence level. For generation \emph{attack vector}, the ECE and BS are 0.87 and 1.00 respectively rising from 0.47 and 0.43. 

\begin{table}
\caption{ECE and BS with \textit{gpt4o} for information generation and extraction.}
\label{tab:calibration_data_extraction}
\resizebox{\columnwidth}{!}{%
\begin{tabular}{cc}
\begin{tabular}{lrrrrrr}
\toprule
& \multicolumn{6}{c}{\textbf{Information extraction}} \\ \midrule
& \multicolumn{2}{c}{\textbf{zero-shot}} & \multicolumn{2}{c}{\textbf{few-shot}} & \multicolumn{2}{c}{\textbf{fine-tuning}} \\
& \textbf{ECE}      &{\textbf{BS}} & \textbf{ECE}  &\textbf{BS} & \textbf{ECE}  & \textbf{BS} \\
\midrule
campaign & 0.25     &{0.26} & 0.26 &{0.28} & 0.48 & 0.48 \\ 
APT & 0.16     &{0.15} & 0.17 &{0.15} & 0.25 & 0.23 \\ 
CVE & 0.28     &{0.32} & 0.35 &{0.37} & 0.18 & 0.21 \\
attack vector & 0.13     &{0.46} & 0.19 &{0.49} & 0.27 & 0.58 \\ 
\bottomrule
\end{tabular}
~
\begin{tabular}{lrrrrrr}
\toprule
&\multicolumn{6}{c}{\textbf{Information generation}} \\ \midrule
& \multicolumn{2}{c}{\textbf{zero-shot}} & \multicolumn{2}{c}{\textbf{few-shot}} & \multicolumn{2}{c}{\textbf{fine-tuning}} \\
& \textbf{ECE}      &{\textbf{BS}} & \textbf{ECE}  &{\textbf{BS}} & \textbf{ECE}  & \textbf{BS}   \\ \midrule
goals & 0.13     &{0.14} & 0.04 &{0.03} & 0.08 & 0.05 \\
labels & 0.45     &{0.44} & 0.57 &{0.53} & 0.48 & 0.49 \\
country & 0.19     &{0.22} & 0.38 &{0.27} & 0.35 & 0.29 \\
CVE & 0.15     &{0.29} & 0.13 &{0.22} & 0.91 & 0.98 \\ 
attack vector & 0.47     &{0.43} & 0.48 &{0.42} & 0.87 & 1.00 \\ \bottomrule
\end{tabular}
\end{tabular}
}
\end{table}

\section{Discussion}
Regarding information extraction, we could see how for \emph{campaign} entity the recall of each model is below 80\%, meaning that the \emph{campaign} entity is in more than 20\% of the cases overlooked. Regarding the \emph{CVE} codes we could see how the maximum recall is 0.90 (\textit{gemini} and \textit{mistral}) and the minimum is 0.87 (\textit{gpt4o}), meaning that LLMs overlook at least the 10\% of the vulnerabilities contained in the threat report and exploited during threat campaigns. The same phenomena concern \emph{attack vector}, in which the maximum recall is 0.83 and the minimum is 0.74, meaning that 26\% of the techniques used to exploit CVEs are overlooked. Few-shot learning and fine-tuning do not improve LLM performance and present detrimental effects after their use, as can be seen in Table \ref{tab:table_extraction_rq1} focusing on the maximum decrease brought by few-shot learning and fine-tuning of respectively 0.07 and 0.19 related to the \emph{APT}. 

The same pattern can be seen in information generation, when the LLM is used to recover the APT profile starting from its name and description, allowing to gather strategic and tactical information. The lowest recall and precision registered are 0.02 when the LLM is used to generate the label of the APT, meaning that the lack of knowledge regarding the type of APT (e.g., nation-state actor, criminal organization, etc.) is almost complete. Low and precision are also present when dealing with \emph{CVE} entity as the minimum precision and recall are 0.10 and 0.06 (\textit{gpt4o}) and the maximum precision and recall are 0.21 and 0.17 (\textit{mistral}). With \emph{country} entity the recall oscillates between 0.70 (\textit{gpt4o}) and 0.78 (\textit{mistral}), meaning that the best LLM in the 20\% of the cases wrongly derives the country of an APT, leading to errors in the attribution of the cyber attacks. As for information extraction, the performance is rarely improved by the application of few-shot learning and fine-tuning, and their use can also be detrimental as shown by Table \ref{tab:table_generation_rq1} (in the Appendix), where in the worst case few-shot learning decreases precision and recall of 0.14 (\emph{labels} entity) and fine-tuning of 0.21 for precision and 0.17 for recall (\emph{CVE} entity) leading to recall and precision of 0.00 when \textit{mistral} generates CVEs exploited by APTs. 

Our analysis highlights the (in)consistency of LLMs outputs. The repeated iteration and bootstrapping techniques are effective in building performance CIs that quantify LLM consistency. The computed CI show that in the task of information extraction from unstructured threat reports, LLMs lack complete determinism, with a maximum CI width of 0.02 (Table \ref{tab:intervals_node_extraction}), posing an additional burden to their use in real-world environments, where repeated analysis could generate contradictory answers regarding the same threat report. Even larger lack of consistency, involves information generation, with a maximum CI width of 0.06, posing limitations to the use of LLMs as CTI chatbots and assistants.

Finally, our work shows that LLMs are not calibrated for CTI as shown by ECE and BSs. Few-shot learning and fine-tuning do not improve model calibration and can also be detrimental. The lack of calibration poses a third burden to using LLMs in real-world scenarios, where a labelled dataset is not available and thus the parameter used to accept or reject the model prediction consists of the confidence assigned to the information extracted or generated. 

\section{Limitations} 
Regarding the performance evaluation, a limitation is posed by evaluating LLMs on only one dataset, as evaluating LLMs on more intelligence reports would allow us to gain a deeper comprehension of the limitations posed by different types of CTI reports. We tried to handle this limitation by selecting a dataset characterized by great heterogeneity in terms of length and report source to reproduce a real-world scenario on which to test LLMs. Another limitation is posed by the limited number of LLMs, which we mitigated by choosing three state-of-the-art LLMs characterized by different architecture and from different providers. 

Second, we consider limitations of the consistency quantification, where we highlight the cost-precision trade-off, as the higher the number of re-prompting the higher the precision and the higher the computational and economical costs. As we quantify the LLM consistency by re-prompting them ten times there is a risk of obtaining unreliable confidence intervals due to bootstrapping with a relatively small sample size. Raising the number of re-prompting would cause higher costs which would lead to the impossibility of reproducing the experiments. However, since the distribution of the samples is unknown, this is, to the best of our knowledge, the most appropriate method. We perform the calibration analysis only with \textit{gpt4o} as it is the only closed-source model between the ones used that allows extracting log probabilities and thus calculating the confidence level assigned by the LLM to tokens. This characteristic of closed-source models also limits the application of post-processing calibration methods such as Platt Scaling which requires access to logits \cite{guo2017calibration,platt1999probabilistic}. We chose closed-source models as they can be run directly on the cloud of proprietary companies, simplifying experiment reproducibility and reducing computational time and economic cost as open-source models need powerful and expensive hardware to be run  \cite{fieblinger2024actionable}.

\section{Conclusion and future work}
We show that LLMs are not ready for real-world CTI tasks. Regarding information extraction, performed on a dataset of real-size reports, given the low precision and recall, the performance cannot guarantee a faithful reconstruction of an attack scenario. Few-shot learning and fine-tuning does not seem to help. We observe the same pattern for information generation, used to build the profile of an APT, indicating that it would be risky to use LLMs as CTI assistants. 
A further worrying aspect consists in the performance oscillations measured and in the lack of calibration which further limits the trustworthiness posed in LLM predictions in a context where few evaluation datasets are available thus model confidence is a fundamental factor in choosing whether to rely on or not on model predictions. 

In future work we plan to extend our experiments to other datasets, such as vulnerability databases, and to more LLMs, to generalize our results regarding the performance limitations that characterize them. Moreover, we plan to experiment with other prompting techniques such as Chain-of-Thought (CoT) \cite{wei2022chain}, with other AI frameworks such as Retrieval Augmented Generation (RAG) \cite{lewis2020retrieval} to improve information generation, and with approaches involving more than one LLM contemporarily, such as LLMs-based multi-agent systems \cite{guo2024large}.
Regarding measuring the consistency of LLMs, we plan to improve the empirical analysis by relying on the combination of different consistency quantification methods \cite{xiong2024efficient} gaining more insights regarding the model's consistency. At the same time, we also plan to integrate the empirical work with a formal analysis, thus complementing the empirical investigation. 

\subsubsection*{Acknowledgements.}
This work was partially supported by the \textit{Nederlandse Organisatie voor Wetenschappelijk Onderzoek (NWO)} under the KIC HEWSTI Project under grant no. KIC1.VE01.20.004, and the Horizon Europe Sec4AI4Sec Project under grant no. 101120393.

\subsubsection*{CRediT Author Statement.} Conceptualization: EM, FM, KT; Methodology: EM; Software: EM; Validation: EM; Formal analysis: na; Investigation: EM; Resources: KT, FM; Data Curation: FM; Writing - Original Draft: EM, FM, KT; Writing - Review \& Editing: FM, KT; Visualization: EM; Supervision: KT, FM; Project administration: FM, KT; Funding acquisition: KT, FM.

\bibliographystyle{unsrt}  

\appendix

\newpage
\section{Additional tables}

\begin{table}[h]
\caption{Results for information generation. Few-shot learning and fine-tuning have limited performance improvement and their effect can be detrimental, as we can see for the entities \emph{country} and \emph{labels}. When a fine-tuned model the performance considerably decreases, even under the performance obtained with zero-shot learning, as it happens with \emph{CVE}, and \emph{labels}.}
\label{tab:table_generation_rq1}
\begin{center}
\resizebox{0.70\columnwidth}{!}{%
\begin{tabular}{llrrrrrrrrr}
\toprule
 &&
  \multicolumn{3}{c}{\textbf{zero-shot}} &
  \multicolumn{3}{c}{\textbf{few-shot}} &
  \multicolumn{3}{c}{\textbf{fine-tuning}} \\
 & \textbf{Model} &
  \textbf{P} &
  \textbf{R} &
 {\textbf{F1}} &
  \textbf{P} &
  \textbf{R} &
 {\textbf{F1}} &
  \textbf{P} &
  \textbf{R} &
  \textbf{F1} \\ \midrule
goals
 &
gpt4o &
  0.85 &
  0.85 &
 {0.85} &
  0.96 &
  0.96 &
 {0.96} &
  0.96 &
  0.96 &
  0.96 \\
&gemini &
  0.72 &
  0.72 &
 {0.72} &
  0.83 &
  0.83 &
 {0.83} &
  0.84 &
  0.84 &
  0.84 \\
&mistral &
  0.77 &
  0.77 &
 {0.77} &
  0.92 &
  0.92 &
 {0.92} &
  0.96 &
  0.96 &
  0.96 \\ \cmidrule{2-11}
labels &
gpt4o &
  0.50 &
  0.50 &
 {0.50} &
  0.44 &
  0.44 &
 {0.44} &
  0.44 &
  0.44 &
  0.44 \\
&gemini &
  0.02 &
  0.02 &
 {0.02} &
  0.54 &
  0.54 &
 {0.54} &
  0.40 &
  0.40 &
  0.40 \\
&mistral &
  0.02 &
  0.02 &
 {0.02} &
  0.36 &
  0.36 &
 {0.36} &
  0.44 &
  0.44 &
  0.44 \\ \cmidrule{2-11}
country &
gpt4o &
  0.70 &
  0.70 &
 {0.70} &
  0.56 &
  0.56 &
 {0.56} &
  0.60 &
  0.60 &
  0.60 \\
&gemini &
  0.73 &
  0.73 &
 {0.73} &
  0.81 &
  0.71 &
 {0.76} &
  0.68 &
  0.68 &
  0.68 \\
&mistral &
  0.78 &
  0.78 &
 {0.78} &
  0.64 &
  0.64 &
 {0.64} &
  0.60 &
  0.60 &
  0.60 \\ \cmidrule{2-11}
CVE &
gpt4o &
  0.10 &
  0.06 &
 {0.08} &
  0.08 &
  0.07 &
 {0.07} &
  0.00 &
  0.00 &
  0.00 \\
&gemini &
  0.13 &
  0.13 &
 {0.13} &
  0.23 &
  0.19 &
 {0.21} &
  0.17 &
  0.36 &
  0.23 \\
&mistral &
  0.21 &
  0.17 &
 {0.19} &
  0.24 &
  0.24 &
 {0.24} &
  0.00 &
  0.00 &
  0.00 \\ \cmidrule{2-11}
attack vector &
gpt4o &
  0.37 &
  0.52 &
 {0.43} &
  0.37 &
  0.51 &
 {0.43} &
  1.00 &
  0.09 &
  0.16 \\
&gemini &
  0.24 &
  0.54 &
 {0.33} &
  0.27 &
  0.56 &
 {0.36} &
  0.52 &
  0.84 &
  0.64 \\
&mistral &
  0.22 &
  0.58 &
 {0.32} &
  0.20 &
  0.75 &
 {0.32} &
  1.00 &
  0.09 &
  0.16 \\ \bottomrule
\end{tabular}}
\end{center}
\,\vspace*{-\baselineskip}
\end{table}

\begin{table}
\,\vspace*{-2\baselineskip}
\caption{Performance CI for the task of information generation. \vspace*{-\baselineskip}}
\label{tab:intervals_node_generation}
\begin{center}
\resizebox{0.90\columnwidth}{!}{%
\begin{tabular}{llcccccc}
\toprule
    && \multicolumn{3}{c}{\textbf{few-shot}} & \multicolumn{3}{c}{\textbf{fine-tuning}} \\
    & \textbf{Models} &\textbf{P}  &\textbf{R} &\textbf{F1} &\textbf{P} &\textbf{R} &\textbf{F1} \\
\midrule
goals                                              
&gpt4o & {[}0.96, 0.96{]} & {[}0.96, 0.96{]} &{{[}0.96, 0.96{]}} & {[}0.96, 0.96{]} & {[}0.96, 0.96{]} & {[}0.96, 0.96{]} \\
&gemini     & {[}0.87, 0.90{]} & {[}0.87, 0.90{]} &{{[}0.87, 0.90{]}} & {[}0.84, 0.84{]} & {[}0.84, 0.84{]} & {[}0.84, 0.84{]} \\
&mistral    & {[}0.92, 0.92{]} & {[}0.92, 0.92{]} &{{[}0.92, 0.92{]}} & {[}0.96, 0.96{]} & {[}0.96, 0.96{]} & {[}0.96, 0.96{]}  \\ \cmidrule{2-8}
labels                                                                                               
&gpt4o & {[}0.44, 0.44{]} & {[}0.44, 0.44{]} &{{[}0.44, 0.44{]}} & {[}0.44, 0.44{]} & {[}0.44, 0.44{]} & {[}0.44, 0.44{]} \\
&gemini     & {[}0.54, 0.56{]} & {[}0.54, 0.56{]} &{{[}0.54, 0.56{]}} & {[}0.39, 0.40{]} & {[}0.39, 0.40{]} & {[}0.39, 0.40{]} \\
&mistral    & {[}0.36, 0.36{]} & {[}0.36, 0.36{]} &{{[}0.36, 0.36{]}} & {[}0.44, 0.44{]} & {[}0.44, 0.44{]} & {[}0.44, 0.44{]} \\ \cmidrule{2-8}
country                                                                                                 
&gpt4o & {[}0.57, 0.59{]} & {[}0.57, 0.59{]} &{{[}0.57, 0.59{]}} & {[}0.60, 0.61{]} & {[}0.60, 0.61{]} & {[}0.60, 0.61{]} \\
&gemini     & {[}0.82, 0.86{]} & {[}0.71, 0.74{]} &{{[}0.76, 0.79{]}} & {[}0.68, 0.68{]} & {[}0.68, 0.68{]} & {[}0.68, 0.68{]} \\
&mistral    & {[}0.64, 0.64{]} & {[}0.64, 0.64{]} &{{[}0.64, 0.64{]}} & {[}0.60, 0.61{]} & {[}0.60, 0.61{]} & {[}0.60, 0.61{]} \\ \cmidrule{2-8}
CVE                                                                                                     
&gpt4o & {[}0.08, 0.09{]} & {[}0.07, 0.08{]} &{{[}0.07, 0.08{]}} & {[}0.00, 0.00{]} & {[}0.00, 0.00{]} & {[}0.00, 0.00{]} \\
&gemini     & {[}0.21, 0.26{]} & {[}0.19, 0.25{]} &{{[}0.20, 0.25{]}} & {[}0.16, 0.19{]} & {[}0.32, 0.37{]} & {[}0.21, 0.24{]} \\
&mistral    & {[}0.23, 0.24{]} & {[}0.24, 0.24{]} &{{[}0.23, 0.24{]}} & {[}0.00, 0.00{]} & {[}0.00, 0.00{]} & {[}0.00, 0.00{]} \\ \cmidrule{2-8}
attack vector                                                                                       
&gpt4o & {[}0.37, 0.37{]} & {[}0.51, 0.52{]} &{{[}0.43, 0.44{]}} & {[}1.00, 1.00{]} & {[}0.09, 0.10{]} & {[}0.16, 0.18{]} \\
&gemini     & {[}0.24, 0.27{]} & {[}0.55, 0.58{]} &{{[}0.34, 0.36{]}} & {[}0.49, 0.50{]} & {[}0.82, 0.84{]} & {[}0.61, 0.63{]} \\
&mistral    & {[}0.20, 0.20{]} & {[}0.74, 0.75{]} &{{[}0.31, 0.31{]}} & {[}1.00, 1.00{]} & {[}0.00, 0.00{]} & {[}0.00, 0.00{]} \\ \bottomrule
\end{tabular}}
\end{center}
\end{table}

\end{document}